\documentclass{stacs_proc}

\RequirePackage{amssymb,amsmath}

\begin{document}

\title[Weak index versus Borel rank]{Weak index versus Borel rank}
\author[lab1]{Filip Murlak}{Filip Murlak}
\address[lab1]{Warsaw University}
\email{fmurlak@mimuw.edu.pl}
\thanks{Supported by the Polish government grant no. N206 008 32/0810.}

\begin{abstract}
  \noindent We investigate weak recognizability of deterministic languages of infinite trees. We prove that for deterministic languages the Borel hierarchy and the weak index hierarchy coincide. Furthermore, we propose a procedure computing for a deterministic automaton an equivalent minimal index weak automaton with a quadratic number of states. The algorithm works within the time of solving the emptiness problem. 
\end{abstract}

\maketitle

\stacsheading{2008}{573-584}{Bordeaux}
\firstpageno{573}

\section{Introduction}

Finite automata on infinite trees are one of the basic tools in the verification of non-terminating programs. Practical applicability of this approach relies on the simplicity of the automata used to express the specifications. On the other hand it is convenient to write the specifications in an expressive language, e.~g.~$\mu$-calculus. This motivates the search for automatic simplifications of automata. An efficient, yet reasonably expressive, model is offered by weak alternating automata. It was essentially showed by Rabin \cite{rabin1} that a language $L$ can be recognized by a weak automaton if and only if both $L$ and $L^\complement$ can be recognized by nondeterministic B\"uchi automata. Arnold and Niwi\'nski \cite{an} proposed an algorithm that, given two B\"uchi automata recognizing a language and its complement, constructs a doubly exponential alternation free $\mu$-calculus formula defining $L$, which essentially provides an equally effective translation to a weak automaton. Kupferman and Vardi \cite{kv} gave an immensely improved construction that involves only quadratic blow-up. 

A more refined construction could also simplify an automaton in terms of different complexity measures. A measure that is particularly important for theoretical and practical reasons is the Mostowski--Rabin index. This measure reflects the alternation depth of positive and negative events in the behaviour of a verified system. The index orders automata into a hierarchy that was proved strict for deterministic \cite{wagner}, nondeterministic \cite{klony}, alternating \cite{brad,lenzi}, and weak alternating automata \cite{most}. Computing the least possible index for a given automaton is called the index problem. Unlike for $\omega$-words, where the solution was essentially given already by Wagner \cite{wagner}, for trees this problem in its general form remains unsolved. For deterministic languages,  Niwi\'nski and Walukiewicz gave algorithms to compute the deterministic and nondeterministic indices \cite{kwiatek,hie}. 

The theoretical significance of the weak index is best reflected by its coincidence with the quantifier alternation depth in the weak monadic second order logic \cite{most}. Further interesting facts are revealed by the comparison with the Borel rank. In 1993 Skurczy\'nski gave examples of $\Pi^0_n$ and $\Sigma^0_n$-complete languages recognized by weak alternating automata with index $(0,n)$ and $(1,n+1)$ accordingly \cite{skurcz}. In \cite{wata} it was shown that weak $(0,n)$-automata can only recognize $\Pi^0_n$ languages (and dually, $(1,n+1)$-automata can only recognize $\Sigma^0_n$ languages), and it was conjectured that the weak index and the Borel hierarchies actually coincide. Here we prove that the conjecture holds for deterministic languages. Consequently, the algorithm calculating the Borel rank for deterministic languages \cite{split} can be also used to compute the weak index. Since all deterministic languages are at the first level of the alternating hierarchy, this completes the picture for the deterministic case. We also provide an effective translation to a weak automaton with a quadratic number of states and the minimal index.

\section{Automata}

We will be working with deterministic and weak automata, but to have a uniform framework, we first define automata in their most general alternating form.

A {\em parity game} is a perfect information game of possibly infinite duration played by two players, Adam and Eve. We present it as a tuple $(V_\exists, V_\forall, E, v_0,\mathrm{rank})$, where $V_\exists $ and $V_\forall$ are (disjoint) sets of positions of Eve and Adam, respectively, $E \subseteq V \times  V $ is the relation of possible moves, with $V =  V_\exists \cup V_\forall $, $p_0 \in V$ is a designated initial position, and $ \mathrm{rank}  : V \to \{0,1, \ldots, n\}$ is the ranking function.

The players start a play in the position $v_0$ and then move a token according to relation $E$ (always to a successor of the current position), thus forming a path in the graph $(V, E)$. The move is selected by Eve or Adam, depending on who is the owner of the current position. If a player cannot  move, she/he looses. Otherwise, the result of the play is an infinite path in the graph, $v_0,v_1,v_2,\ldots $. Eve wins the play if the highest rank visited infinitely often is even, otherwise Adam wins.

An {\em alternating automaton} $A=\langle \Sigma, Q_\exists, Q_\forall, q_0, \delta, \mathrm{rank}\rangle$, consists of a finite input alphabet $\Sigma$, a finite set of states $Q$ partitioned into existential states $Q_\exists$ and universal states $Q_\forall$ with a fixed initial state $q_0$, a transition relation $\delta \subseteq Q \times \Sigma \times \{0,1, \varepsilon\}\times Q$, and a ranking function $\mathrm{rank}:Q \to \omega$. Instead of $(p,\sigma, d, q)\in \delta$, one usually writes $p\stackrel{\sigma,d}{\longrightarrow}q$.

An input tree $t$ is accepted by $A$ iff Eve has a winning strategy in the parity game  $\langle Q_\exists \times \{0,1\}^*, Q_\forall \times \{0,1\}^*, (q_0,\varepsilon), E, \mathrm{rank'}\rangle$, where $E= \{((p,v),(q,vd))\colon v\in \mathrm{dom}(t),\; (p, t(v), d, q)\in \delta\}$ and $\mathrm{rank'} (q,v) = \mathrm{rank}(q)$. The computation tree of $A$ on $t$ is obtained by unravelling the graph above from the vertex $(q_0, \varepsilon)$ and labelling the node $(q_0,\varepsilon), (q_1, d_1), (q_2, d_2), \ldots, (q_n,d_n)$ with $q_n$. The result of the parity game above only depends on the computation tree. 

An automaton is called {\em deterministic} iff Eve has no choice at all, and Adam can only choose the direction: left or right (no $\varepsilon$-moves). Formally, it means that $Q_\exists=\emptyset$, and $\delta: Q \times \Sigma \times\{0,1\} \to Q$. For deterministic automata, the computation tree is a full binary tree. The transitions are often written as $p\stackrel{\sigma}{\longrightarrow}q_0,q_1$, meaning $p\stackrel{\sigma, d}{\longrightarrow}q_d$ for $d=0,1$.

A {\em weak automaton} is an alternating automaton satisfying the condition \[ p \stackrel{\sigma,d}{\longrightarrow} q \quad \implies \quad \mathrm{rank}\, p \leq \mathrm{rank}\, q\,.\] A more elegant definition of the class of weakly recognizable languages is obtained by using {\em weak parity games} in the definition of acceptance by alternating automata. In those games Eve wins a play if the highest rank used at least once is even. For the purpose of the following lemma, let us call the first version {\em restricted alternating automata}. Later, we will stick to the second definition.

\begin{lemma} For every $L$ it holds that $L$ is recognized by a restricted alternating $(\iota,\kappa)$-automaton iff it is recognized by a weak alternating $(\iota,\kappa)$-automaton.
\end{lemma}
\proof Every restricted automaton can be transformed into an equivalent weak automaton by simply changing the acceptance condition to weak. Let us, then,  concentrate on the converse implication. 

Fix a weak automaton $A$ using ranks $(\iota, \kappa)$. To construct a restricted automaton we will take one copy of $A$ for each rank: $A^{(\iota)}, A^{(\iota+1)}, \ldots, A^{(\kappa)}$. By $q^{(i)}$ we will denote the counterpart of $A$'s state $q$ in $A^{(i)}$. We set $\mathrm{rank}\, q^{(i)} = i$.  We want the number of the copy the computation is in to reflect the highest rank seen so far. To obtain that, we set the initial state of the new automaton to $q_0^{(\mathrm{rank}\, q_0)}$, and for each $i$ and each transition $p\stackrel{\sigma, d}{\longrightarrow}q$ in $A$ we add a transition $p^{(i)}\stackrel{\sigma, d}{\longrightarrow}q^{(\max(i,\, \mathrm{rank}\, q))}$. For each $i$ and $q$, $q^{(i)}$ is universal iff $q$ is universal. Checking the equivalence is straightforward. \qed

For deterministic automata we will assume that all states are productive, i.~e., are used in some accepting run, save for one all-rejecting state $\bot$, and that all transitions are productive or go to $\bot$, i.~e., whenever  $q\stackrel{\sigma}{\longrightarrow} q_1, q_2$, then either $q_1$ and $q_2$ are productive, or $q_1=q_2=\bot$. The assumption of productivity is vital for our proofs. Thanks to this assumption, in each node of an automaton's run we can plug in an accepting sub-run.

Transforming a given automaton into such a form of course needs calculating the productive states, which is equivalent to deciding a language's emptiness. The latter problem is known to be in $\textrm{NP} \cap \textrm{co-NP}$, but it has no polynomial solutions yet. Therefore we can only claim that our algorithms are polynomial for the automata that underwent the above preprocessing. We will try to mention it whenever particularly important.

\section{Two Hierarchies}

The {\em index of an automaton} $A$ is a pair $(\min {\rm rank}\, Q,\max {\rm rank}\, Q )$. Scaling down the rank function if necessary, one may assume that $\min {\rm rank}\, Q$ is either 0 or 1. Thus, the indices are elements of $\{0,1\}\times\omega \setminus \{(1,0)\}$. For an index $(\iota,\kappa)$ we shall denote by $\overline {(\iota,\kappa)}$ {\em the dual index}, i.~e., $\overline{(0,\kappa)} = (1,\kappa+1)$, $\overline{(1,\kappa)} = (0,\kappa-1)$. Let us define an ordering of indices with the following formula: \[(\iota,\kappa) < (\iota',\kappa')  \textrm{ if and only if } \kappa - \iota < \kappa' < \iota' \,.\] In other words, one index is greater than another if and only if it ``uses'' more ranks. This means that dual indices are incomparable.  
 {\em The Mostowski--Rabin index hierarchy} for a certain class of automata consists of ascending sets (levels) of languages recognized by $(\iota,\kappa)$-automata.

\begin{figure}
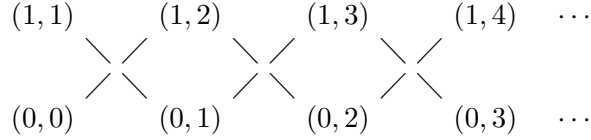

\centering
{\setlength\arraycolsep{2pt}
$\begin{array}{cccccccccccccc}
(1,1) &&& (1,2) &&& (1,3) &&& (1,4)  & \quad \cdots\\
& \diagdown & \diagup && \diagdown & \diagup && \diagdown & \diagup \\ 
& \diagup & \diagdown && \diagup & \diagdown && \diagup & \diagdown \\  
(0,0) &&& (0,1) &&& (0,2) &&& (0,3) & \quad \cdots 
\end{array}$}
\caption{The Mostowski--Rabin index hierarchy}
\end{figure}

Here, we are mainly interested in the {\em weak index hierarchy}, i.~e., the hierarchy of languages recognized by weak $(\iota,\kappa)$-automata. The strictness of this hierarchy was established by Mostowski \cite{most} via equivalence with the quantifier-alternation hierarchy for the weak monadic second order logic, whose strictness was proved by Thomas \cite{weakthomas}. The {\em weak index problem}, i.~e., computing the minimal weak index needed to recognize a given weak language, for the time being remains unsolved just like other versions of the index problem. 

The weak index hierarchy is closely related to the Borel hierarchy. We will work with the standard Cantor-like topology on $T_{\Sigma}$ induced by the metric 
\[ d(s,t) = \left \{ 
  \begin{array} {l l}
    2 ^{-\min \{|x|\;: \;\; x \in \{0,1\}^*, \; s(x) \neq t(x)\}} & \textrm{iff } s \neq t \\
    0 & \textrm{iff } s=t
  \end{array}
\right . .\]
The class of Borel sets of a topological space $X$ is the closure of the class of open sets of $X$ by countable sums and complementation.  

\begin{figure}
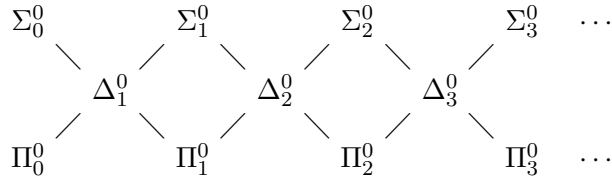

\centering
\vspace{4ex}
{\setlength\arraycolsep{2pt}
$\begin{array}{cccccccccccccc}
\Sigma^0_0 &&&& \Sigma^0_1 &&&& \Sigma^0_2 &&&& \Sigma^0_3  & \quad \cdots\\
& \diagdown && \diagup && \diagdown && \diagup && \diagdown && \diagup \\ 
&& \Delta^0_1 &&&& \Delta^0_2 &&&& \Delta^0_3 \\
& \diagup && \diagdown && \diagup && \diagdown && \diagup && \diagdown \\  
\Pi^0_0 &&&& \Pi^0_1 &&&& \Pi^0_2 &&&& \Pi^0_3  & \quad \cdots
\end{array}$}
\caption{The Borel hierarchy}
\end{figure}

\noindent For a topological space $X$, the initial (finite) levels of the {\em Borel hierarchy} are defined as follows:
\begin{itemize}
  \item $\Sigma^0_1(X)$ --  open subsets of $X$,
  \item $\Pi^0_k(X)$ --  complements of the sets from $\Sigma^0_k(X)$, 
  \item $\Sigma^0_{k+1}(X)$ --  countable unions of sets from $\Pi^0_k(X)$.
\end{itemize}
For instance, $\Pi^0_1(X)$ are the closed sets, $\Sigma^0_2(X)$ are $F_\sigma$ sets and $\Pi^0_2(X)$ are $G_\delta$ sets. By convention $\Sigma^0_0(X) = \{\emptyset\}$ and $\Pi^0_0(X) = \{X\}$. 

A straightforward inductive argument shows that the classes defined above are closed under inverse images of continuous functions. Let ${\mathcal C}$ be one of those classes. A set $A$ is called {\em ${\mathcal C}$-hard}, if each set in ${\mathcal C}$ is an inverse image of $A$ under some continuous function. If additionally $A\in{\mathcal C}$, $A$ is {\em ${\mathcal C}$-complete}. 

We start the discussion of the relations between the index of a weak automaton and the Borel rank of the language it recognises by recalling Skurczy\'nski's results. For a tree $t:\{0,1\}^* \to \Sigma$ and a node $v\in \{0,1\}^*$ let $t.v$ denote the tree rooted in $v$, i. e., $t.v(w) = t(vw)$. Let us define a sequence of languages:

\begin{itemize}
\item $L_{(0,1)} = \{t\}$, where $t \in T_{\{a,b\}}$ is the tree with no $b$'s,
\item $L_{(1,n+1)} = L_{(0,n)} ^\complement$ for $n \geq 1$,
\item $L_{(0,n+1)} = \{t \in T_{\{a,b\}}: \;\; \forall_{k} \; t.0^k1 \in L_{(1,n+1)}\}$ for $n \geq 1$. 
\end{itemize}

\begin{theorem}[Skurczy\'nski \cite{skurcz}]
For each $n \geq 1$,  
\begin{itemize}
\item $L_{(0,n)}$ is a $\Pi^0_n$-complete language recognized by a weak $(0,n)$-automaton,
\item $L_{(1,n+1)}$ is a $\Sigma^0_n$-complete language recognized by a weak $(1,n+1)$-automaton.
\end{itemize}
\end{theorem}

\noindent We will now show that this construction is as efficient as it can be: ranks $(0,n)$ are necessary to recognize any $\Pi^0_n$-hard language (if it can be weakly recognized at all).

We will actually prove a~bit stronger result. We will consider {\em weak game languages} $W_{[\iota, \kappa]}$, to which all languages recognized by weak $[\iota, \kappa]$-automata can be reduced, and show that $W_{[0,n]} \in \Pi^0_n$ and $W_{[1,n+1]} \in \Sigma^0_n$ (by Skurczy\'nski's results, they are hard for these classes). The languages $W_{[\iota, \kappa]}$ are natural weak counterparts of strong game languages that prove the strictness of the strong alternating index hierarchy. Lately Arnold and Niwi\'nski proved that the strong game languages also form a~strict hierarchy with respect to continuous reductions, but they are all non-Borel \cite{gamelang}.

Fix a natural number $N$. For $\iota=0,1$ and $\kappa \geq \iota$,
let ${\mathcal T}_{(\iota, \kappa)}$ denote the set of full $N$-ary trees over the alphabet $\{\exists,\forall\} \times \{\iota, \iota+1, \ldots, \kappa\}$. Let $W_{(\iota, \kappa)} \subseteq {\mathcal T}_{(\iota, \kappa)}$ be the set of all trees $t$ for which Eve has a~winning strategy in the {\em weak} parity game $G_t=\langle V_\exists, V_\forall, E, v_0, \mathrm{rank} \rangle$, where $V_\theta = \{v \in \mathrm{dom}\, t \colon t(v)=(\theta, j) \textrm{ for some } j\}$, $E= \{ (v,vk)\colon v\in \mathrm{dom}\, t \,,\;  k < N\}$, $v_0 = \varepsilon$, $\mathrm{rank}(v)=j$ iff $t(v) = (\theta,j)$ for some $\theta$.

\begin{theorem} \label{weakgamelang}
For each $n$, $W_{(0,n)}\in \Pi^0_n(\mathcal T_{(0,n)})$ and $W_{(1,n+1)} \in \Sigma^0_n(\mathcal T_{(1,n+1)})$.
\end{theorem}

\proof We will proceed by induction on $n$. For $n=0$ the claim is obvious: $W_{(0,0)} = {\mathcal T}_{(0, 0)} \in \Pi^0_0(\mathcal T_{(0,0)})$, $W_{(1,1)} = \emptyset \in \Sigma^0_0(\mathcal T_{(1,1)})$. 

Take $n>0$. For each $t\in W_{(1,n+1)}$ there exists a strategy $\sigma$ for Eve, such that it guarantees that the play reaches a node with the rank greater or equal to $2$. By K\"onig lemma, this must happen in a bounded number of moves. Basing on this observation we will provide a $\Sigma^0_n$ presentation of $W_{(1,n+1)}$.

Let {\em $k$-antichain} be a subset of the nodes on the level $k$. Let ${\mathcal A}$ denote the set of all possible $k$-antichains for all $k<\omega$. Obviously this set is countable. For a $k$-antichain $A$ let $W_A$ denote the set of trees such that there exists a strategy for Eve that guarantees visiting a node with the rank $\geq 2$ during the initial $k$ moves and reaching a node from $A$. This set is a clopen.  We have a presentation \[ W_{(1,n+1)} =  \bigcup_{A \in {\mathcal A}} \left ( W_A \cap \bigcap_{v \in A} \left \{t: t'.v \in W_{(0,n-1)} \right \} \right ) \,,\] where $t'$ is obtained from $t$ by decreasing all the ranks by $2$ (if the result is $-1$, take $0$). The claim follows by induction hypothesis and the continuity of $t \mapsto t'$ and $t \mapsto t.v$. 

Now, it remains to see that $W_{(0,n)}\in\Pi^0_n(\mathcal T_{(0,n)})$. For this, note that \[W_{(0,n)} = \left \{t: t'' \in (W_{(1,n+1)}) ^\complement \right \}\,,\] where $t''$ is obtained from $t$ by swapping $\exists$ and $\forall$, and increasing ranks by 1. The claim follows by the continuity of $t\mapsto t''$. \qed

\vspace{2mm}

As a~corollary we get the promised improvement of Skurczy\'nski's result.

\begin{corollary} \label{weakborel}
For every weak alternating automaton $A$ with index $(0,n)$ (resp. $(1,n+1)$) it holds that $L(A)\in \Pi^0_n$ (resp. $L(A)\in \Sigma^0_n$).
\end{corollary}

\proof Let $A$ be an automaton with priorities inside $[\iota, \kappa]$.  For sufficiently large $N$ we may assume without loss of generality that the computation trees of the automaton are $N$-ary trees. By assigning to an input tree the run of $A$, one obtains a~continuous function reducing $L(A)$ to $W_{(\iota, \kappa)}$. Hence, the claim follows from the theorem above. \qed

\vspace{2mm}

In fact the corollary follows also from Mostowski's theorem on equivalence of weak automata and weak monadic second order logic on trees \cite{most}. The present proof of Theorem \ref{weakgamelang} is actually just a repetition of Mostowski's proof in the setting of the Borel hierarchy. An entirely different proof can be found in \cite{wata}.

We believe that the converse implication is also true: a weakly recognizable $\Pi^0_n$-language can be recognized by a weak $(0,n)$-automaton (and dually for $\Sigma^0_n$). 
\begin{conjecture} 
For weakly recognizable languages the weak index hierarchy and the Borel hierarchy coincide.
\end{conjecture}
\noindent In this paper we show that the conjecture holds true when restricted to deterministic languages.

\section{The Deterministic Case}

In 2002 Niwi\'nski and Walukiewicz discovered a surprising dichotomy in the family of deterministic languages: a deterministic language is either very simple or very sophisticated. 

\begin{theorem}[Niwi{\'n}ski, Walukiewicz \cite{gap}] \label{topgap}
For a deterministic automaton $A$ with $n$ states, $L(A)$ is either recognizable with a weak alternating $(0,3)$-automaton with ${\mathcal O}(n^2)$ states (and so  $\Pi^0_3$) or is non-Borel (and so not weakly recognizable). 
The equivalent weak automaton can be constructed within the time of solving the emptiness problem.  
\end{theorem}

\begin{figure}
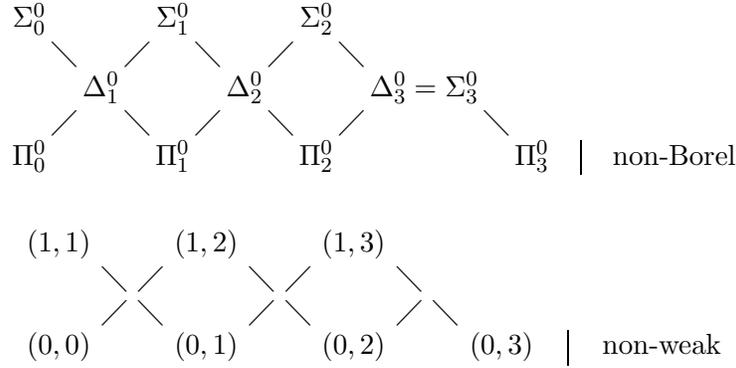

\centering
{\setlength\arraycolsep{1pt}
$\begin{array}{ccccccccccccc|c}
\Sigma^0_0 &&&& \Sigma^0_1 &&&& \Sigma^0_2 \\
& \diagdown && \diagup && \diagdown && \diagup && \diagdown &&\\ 
& & \Delta^0_1 & & & & \Delta^0_2 & & & & \Delta^0_3 = \Sigma^0_3\\
& \diagup &&  \diagdown && \diagup && \diagdown && \diagup & &\diagdown \\  
\Pi^0_0 &&&& \Pi^0_1 &&&& \Pi^0_2 &&&& \Pi^0_3 \quad & \quad \textrm{non-Borel}
\end{array}$}

\vspace{4ex}

{\setlength\arraycolsep{2pt}
$\begin{array}{cccccccccc|c}
(1,1) &&& (1,2) &&& (1,3) \\
& \diagdown & \diagup && \diagdown & \diagup && \diagdown &\\ 
& \diagup & \diagdown && \diagup & \diagdown && \diagup & \diagdown \\  
(0,0) &&& (0,1) &&& (0,2) &&& (0,3) \quad & \quad \textrm{non-weak}
\end{array}$}
\caption{The Borel hierarchy and weak index hierarchy for deterministic tree languages.}
\label{fig:weakindexhierarchy}
\end{figure}

An important tool used in the proof of the Gap Theorem (Theorem \ref{topgap}) is the technique of difficult patterns. In the topological setting the general recipe goes like this: for a given class identify a pattern that can be unravelled to a language complete for this class; if an automaton does not contain the pattern, then $L(A)$ should be in the dual class. The same technique was later applied to obtain effective characterisations of the remaining Borel classes of deterministic languages \cite{split}. 

Let us define the patterns used in these characterisations. A {\em loop} in an automaton is a sequence of states and transitions: \[p_0 \stackrel{\sigma_1,d_1}{\longrightarrow} p_1 \stackrel{\sigma_2,d_2}{\longrightarrow} \ldots \stackrel{\sigma_{n},d_n}{\longrightarrow} p_0\,.\] A loop is called {\em accepting} if $\max_i \mathrm{rank}\,(p_i)$ is even. Otherwise it is {\em rejecting}.  

A {\em $(\iota,\kappa)$-flower} is a sequence of loops $\lambda_\iota, \lambda_{\iota+1}, \ldots , \lambda_\kappa$ starting in the same state $p$, such that the highest rank appearing on $\lambda_i$ has the same parity as $i$ and it is higher than the highest rank on $\lambda_{i-1}$ for $i=\iota, \iota+1, \ldots, \kappa$. 

A {\em weak $(\iota,\kappa)$-flower} is a sequence of loops $\lambda_\iota, \lambda_{\iota+1} \ldots , \lambda_\kappa$ such that $\lambda_{i+1}$ is reachable from $\lambda_i$, and $\lambda_i$ is accepting iff $i$ is even. 

A {\em split} is a pair of loops $p \stackrel{\sigma, 0}{\longrightarrow} p_0 \longrightarrow \ldots \longrightarrow p$ and $p \stackrel{\sigma,1}{\longrightarrow} p_1 \longrightarrow \ldots \longrightarrow p$ such that the highest ranks occurring on them are of different parity and the highest one is odd.

A state $q$ is {\em replicated} by a loop $p \stackrel {\sigma, d_0}\longrightarrow p_0 \longrightarrow \ldots \longrightarrow p $ if there exists a path $p \stackrel{\sigma, d_1} \longrightarrow p_1 \longrightarrow \ldots \longrightarrow q$ such that $d_0 \neq d_1$. We will say that a loop or a flower is replicated by a loop $\lambda$ if it contains a state replicated by $\lambda$.

\begin{proposition}[Niwi\'nski, Walukiewicz \cite{gap}; Murlak \cite{split}] \label{hardborel}
Let $A$ be a deterministic automaton.
\begin{enumerate}
\item $L(A) \in \Pi^0_1$ iff $A$ contains no weak $(1,2)$-flower.
\item $L(A) \in \Sigma^0_1$ iff $A$ contains no weak $(0,1)$-flower.
\item $L(A)\in\Pi^0_2$ iff $A$ contains no $(0,1)$-flower.
\item  $L(A) \in \Sigma^0_2$ iff $A$ contains neither $(1,2)$-flower nor a weak $(1,2)$-flower replicated by an accepting loop.
\item $L(A)\in \Sigma^0_3$ iff $A$ contains no $(0,1)$-flower replicated by an accepting loop.
\item $L(A)\in\Pi^0_3$ iff $A$ contains no split.
\end{enumerate}
In particular, the Borel rank of $L(A)$ is computable within the time of finding the productive states of $A$.
\end{proposition}

The patterns defined above were originally introduced to capture the index complexity of recognizable languages. Niwi\'nski and Walukiewicz used flowers to solve the deterministic index problem for word languages \cite{kwiatek}. Their result may easily be adapted to trees (see \cite{split} for details). 

\begin{theorem} \label{indch}
For a deterministic tree automaton $A$ the language $L(A)$ is recognized by a deterministic $(\iota,\kappa)$-automaton iff $A$ does not contain a $\overline{(\iota,\kappa)}$-flower. An equivalent minimal index automaton with the same number of states can be constructed within the time of solving the emptiness problem. 
\end{theorem}

\noindent The weak flowers provide an analogous characterisation of the weak deterministic index. 

\begin{proposition}[\cite{split}] \label{windch} 
A deterministic automaton $A$ is equivalent to a weak deterministic $(\iota,\kappa)$-automaton iff it does not contain a weak $\overline{(\iota,\kappa)}$-flower. An equivalent minimal index automaton with the same number of states can be constructed within the time of solving the emptiness problem. 
\end{proposition}

\proof If the automaton contains a weak $(\iota,\kappa)$-flower, for each weak  $\overline{(\iota,\kappa)}$-automaton one can build a cheating tree (see \cite{split} for details). For the converse implication, construct a weak deterministic $(\iota,\kappa)$-automaton by modifying the ranks of the given deterministic automaton. Set ${\rm rank}\,q$ to the lowest number $m$ such that there exists a weak $(m,\kappa)$-flower with a path from $q$ to $\lambda_m$. \qed

\section{The Power of the Weak} \label{sect_weakindex}

In this section we finally turn to the weak recognizability of deterministic languages. First we give sufficient conditions for a deterministic automaton to be equivalent to a weak alternating automaton of index $(0,2)$, $(1,3)$, and $(1,4)$. This is the first step to the solution of the weak index problem for deterministic automata.

\begin{proposition} \label{weak02}
For each deterministic $(1,2)$-automaton with $n$ states one can construct an equivalent weak $(0,2)$-automaton with $2n+1$ states.
\end{proposition}

\proof Fix a deterministic $(1,2)$-automaton $A$. We will construct a weak $(0,2)$-automaton $B$ such that $L(A) = L(B)$. Basically, for each node $v$ the automaton $B$ should check whether on each path in the subtree rooted in $v$ the automaton $A$ will reach a state with rank 2. This can be done as follows. Take two copies of $A$. In the first copy, all states are universal and have rank 0. The transitions are like in $A$ plus for each state $q^{(1)}$ there is an $\varepsilon$-transition to $q^{(2)}$, the counterpart of $q^{(1)}$ in the second copy. In the second copy all states are universal and have rank 1. For the states with rank 1 in $A$, the transitions are like in $A$. For the states with rank 2 in $A$, there is just one transition to an all-accepting state $\top$ (rank 2 in $B$). \qed

Before we proceed with the conditions, let us show a useful property of the replication. 

\begin{lemma}[Replication Lemma] \label{replicationlemma}
A state occurs in infinitely many incomparable nodes of an accepting run iff it is productive and is replicated by an accepting loop. 
\end{lemma}

\proof If a state $p$ is replicated by an accepting loop, then by productivity one may easily construct an accepting run with infinitely many incomparable  occurrences of $p$. Let us concentrate on the converse implication. 

Let $p$ occur in an infinite number of incomparable nodes $v_0, v_1, \ldots$ of an accepting run $\rho$. Let $\pi_i$ be a path of $\rho$ going through the node $v_i$. Since $2^\omega$ is compact, we may assume, passing to a subsequence, that the sequence $\pi_i$ converges to a path $\pi$. Since $v_i$ are incomparable, $v_i$ is not on $\pi$. Let the word $\alpha_i$ be the sequence of states labeling the path from the last common node of $\pi$ and $\pi_i$ to $v_i$. Cutting the loops off if needed, we may assume that $|\alpha_i| \leq |Q|$ for all $i\in \omega$. Consequently, there exist a word $\alpha$ repeating infinitely often in the sequence $\alpha_0, \alpha_1, \ldots$. Moreover, the path $\pi$ is accepting, so the starting state of $\alpha$ must lay on an accepting productive loop. This loop replicates $p$.\qed

\begin{proposition} \label{weak13}
For each deterministic $(0,1)$-automaton with $n$ states which contains no weak $(1,2)$-flower replicated by an accepting loop one can construct effectively an equivalent weak $(1,3)$-automaton with $3n+1$ states.
\end{proposition}

\proof Let $A$ be a deterministic $(0,1)$-automaton which contains no weak $(1,2)$-flower replicated by an accepting loop. Let us call a state of $A$ {\em relevant} if it has the highest rank on some loop. We may change the ranks of productive irrelevant states to $0$, and assume from now on that all odd states are relevant. We claim that the odd states occur only finitely many times on accepting runs of $A$. Suppose that an odd state $p$ occurs infinitely many times in an accepting run $\rho$. Then it must occur in infinitely many incomparable nodes (otherwise we would get a rejecting path). By the Replication Lemma $p$ is replicated by an accepting loop. As $p$ is odd and  relevant, it lies on some nontrivial rejecting loop. Since $p$ is also productive, some accepting loop can be reached from $p$. Hence, $A$ contains a weak $(1,2)$-flower replicated by an accepting loop - a contradiction

Now, we can easily construct a weak $(1,3)$-automaton recognising $L(A)$. Intuitively, we will simulate $A$ and check if $A$'s odd states occur finitely many times. This can be done as follows. Take three copies of $A$. In the first copy all the states are universal and have rank 1. The transitions are just like in $A$, only they go to the second copy of $A$. In the second copy of $A$, all the states are existential and have rank 1. From each state $q^{(2)}$ there are two $\varepsilon$-transitions to $q^{(1)}$ in the first copy and to $q^{(3)}$ in the third copy. Finally, in the third copy of $A$ all the states are universal and have rank 2. The transitions from the states ranked 0 in $A$ are just like in $A$, and from the states ranked $1$ in $A$ they go to an all-rejecting state $\bot$ (rank 3 in $B$). It is easy to see that $B$ recognizes $L(A)$. \qed

\begin{proposition} \label{weak14}
For each automaton with $n$ states containing no $(0,1)$-flower replicated by an accepting loop one can construct an equivalent weak alternating  $(1,4)$-automaton with ${\mathcal O}(n^2)$ states.
\end{proposition}

\proof Let $A$ be an automaton without $(0,1)$-flower replicated by an accepting loop. Consider the DAG of strongly connected components of $A$.For each SCC $X$ containing at least one loop we will construct a weak automaton $B_X$ recognising the languages of trees $t$ such that each path of $A$'s run on $t$ that enters $X$ either leaves $X$ or is accepting. Obviously, the conjunction of such automata recognizes exactly $L(A)$. Let us first consider components replicated by an accepting loop. By the hypothesis, such a component must not contain a $(0,1)$-flower. Therefore we may assume that $X$ only uses ranks 1 and 2. To obtain $B_X$ take a copy of $A$. The states outside $X$ can be divided into three disjoint groups: those that can be reached from $X$, those from which $X$ can be reached, and the rest. Give the states from the first group the rank 4, and the states from the second and third group the rank 2. Finally, following the method from Proposition \ref{weak02}, replace X with an equivalent weak alternating subautomaton using ranks 2,3, and 4. The constructed automaton has ${\mathcal O}(n)$ states.

The case of $X$ not replicated by an accepting loop is more tricky. The key property follows from the Replication Lemma. Let $\rho_X$ denote the restriction of the run $\rho$ to the nodes labeled with a state from $X$ or having a descendant labeled with a state from $X$. By the Replication Lemma, this tree has only finitely many branches (some of them may be infinite).  What $B_X$ should do is to guess a node $v$ on each path such that in the subtree rooted in $v$, $\rho_X$ is either empty or consists of one infinite accepting branch. In the latter case we may additionally demand that on this infinite path the highest rank that ever occurs, occurs infinitely many times.

$B_X$ consists of the component $C_{\textrm{guess}}$ realising the guessing, the component $C_{A \setminus X}$ checking that no path of the computation enters $X$, and components $C_{X,r}$ for all ranks $r$ used in $X$, which check that in a given subtree of the run $\rho$ there is exactly one branch of $\rho_X$ and that on this branch $r$ occurs infinitely often and no higher rank is used.

To construct $C_{\textrm{guess}}$, take a copy of $A$ and declare all the states universal and set their ranks to $1$. For each $q$ add a fresh existential state $q'$ of rank $1$ with an $\varepsilon$-transition to $q$ and either to $q^{A\setminus X}\in C_{A \setminus X}$ if $q\notin X$ ($\rho_X$ is empty) or to $q^{X,r} \in C_{X,r}$ for all $r$ if $q\in X$ ($\rho_X$ is one infinite accepting path). Finally replace each transition $p \stackrel{\sigma}{\longrightarrow}p_0,p_1$ with  $\stackrel{\sigma}{\longrightarrow}p'_0,p'_1$.

The component $C_{A \setminus X}$ is a copy of $A$ with all ranks equal $2$, and the SCC $X$ replaced with one all-rejecting state $\bot$ with rank $3$. 

Finally, let us now describe the automaton $C_{X,r}$. The automaton, staying in rank 2, works its way down the input tree just like $A$ would, with the following modifications:
\begin{itemize}
\item if $A$ enters a state in $X$ with rank greater than $r$, $C_{X,r}$ moves to an all rejecting state $\bot$ (rank 3),
\item if $A$ takes a transition exiting $X$ on both branches or staying in $X$ on both branches, $C_{X,r}$ moves to $\bot$,
\item if $A$ takes a transition whose left branch leaves $X$ and the right branch stays inside, $C_{X,r}$ sends to the right a $(3,4)$-component looking for a state from $X$ with the rank $r$, and moves on to the right subtree (and symmetrically).
\end{itemize}

In order two see that $C_{X,r}$ does the job, it is enough to observe that if the $(3,4)$ component always succeeds to find a state from $X$ with the rank $r$, then on the unique path that stays forever in $X$ the rank $r$ repeats infinitely often. 

The $(3,4)$-component of $C_{X,r}$ can be constructed in such a way that it has $|X|+2$ states, and so in this case $B_X$ has at most $2|X|(|X|+2) + 3n \leq 2|X|^2 + 7n$ states. 

In both cases, the number of states of $B_X$ can be bounded by $c_1|X|^2 + c_2n$ for fixed constants $c_1$ and $c_2$, independent of $X$. Since the SCCs are disjoint, the number of states of the conjunction of $B_X$'s is at most \[ 1 + \sum_{X\in A} (c_1|X|^2 + c_2n) \leq 1+ c_1 \left ( \sum_{X \in A} |X|\right )^2 + c_2n^2 \leq (c_1+c_2)n^2 + 1\,.\]\qed

We have now collected all the ingredients for the solution of the weak index problem for deterministic languages. What is left to be done is to glue together the sufficient conditions for index easiness and Borel hardness using  Corollary \ref{weakborel}.

\begin{figure}
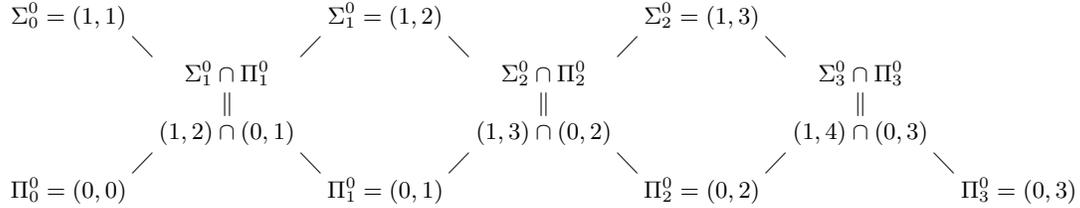

\centering
\footnotesize{
{\setlength\arraycolsep{1pt}
$\begin{array}{ccccccccccccc}
\Sigma^0_0=(1,1) &&&& \Sigma^0_1 = (1,2) &&&& \Sigma^0_2 = (1,3) \\
& \diagdown && \diagup && \diagdown && \diagup && \diagdown &&\\ 
 & & \Sigma^0_1 \cap\Pi^0_1 & & & & \Sigma^0_2 \cap \Pi^0_2 & & & & \Sigma^0_3 \cap \Pi^0_3 \\
 & & \| & & & & \| & & & & \|\\
 & & (1,2) \cap (0,1)& & & & (1,3) \cap (0,2)& & & &  (1,4) \cap (0,3)\\
& \diagup &&  \diagdown && \diagup && \diagdown && \diagup & &\diagdown \\  
\Pi^0_0 = (0,0) &&&& \Pi^0_1 = (0,1) &&&& \Pi^0_2 = (0,2) &&&& \Pi^0_3 = (0,3) \quad 
\end{array}$}}
\caption{For deterministic tree languages the hierarchies coincide.}
\label{fig:dethierarchy} 
\end{figure}

\begin{theorem}
For deterministic languages the Borel hierarchy and the weak index hierarchy coincide (Fig. \ref{fig:dethierarchy}) and are decidable within the time of solving emptiness problem. For a deterministic automaton with $n$ states, an equivalent minimal index automaton with ${\mathcal O}(n^2)$ states can be constructed effectively within the time of solving the emptiness problem.
\end{theorem}

\proof We will abuse the notation and write $(\iota, \kappa)$ to denote the class of languages recognized by weak $(\iota, \kappa)$-automata. All the classes considered here are relativised to the deterministic languages. 

By the two versions of the Gap Theorem we have the equality and decidability of the classes of the classes $\Pi^0_3$ and $(0,3)$. 

Let us continue with the third level. Let us see that $\Sigma^0_3 = (1,4)$. We will show that both these classes are equal to the class of languages recognized by deterministic automata without a $(0,1)$-flower replicated by an accessible loop. If a deterministic automaton $A$ does not contain this pattern, then it is equivalent to a weak $(1,4)$-automaton and by Corollary \ref{weakborel} recognizes a $\Sigma^0_3$ language. If $A$ does contain this pattern, then by Proposition \ref{hardborel} it is not $\Sigma^0_3$ and so is not equivalent to a weak $(1,4)$-automaton. The decidability follows easily, since checking for the pattern above can be done effectively (in polynomial time).

For the equality $\Pi^0_2 = (0,2)$, prove that both classes are equal to the class of languages recognized by deterministic automata without a $(0,1)$-flower. Proceed just like before, only use Proposition  \ref{weak02} instead of Proposition \ref{weak14}. Analogously, using Proposition \ref{weak13}, show that both $\Sigma^0_2$ and $(1,3)$ are equal to the class of languages recognized by deterministic automata admitting neither a $(1,2)$-flower nor a weak $(1,0)$-flower replicated by an accepting loop.

For the first level use the characterisation given by Proposition \ref{windch}. The level zero is trivial. \qed

\section*{Acknowledgments}

The author thanks Damian Niwi\'nski for reading carefully a preliminary version of this paper and the anonymous referees for their helpful comments.

\end{document}